\documentclass[twoside, a4paper, 12pt] {elsart}
\usepackage{graphicx}

\usepackage{amsmath}

\newtheorem{teorema}{Theorem}

\def\CaixaPreta{\vrule Depth0pt height5pt width5pt}
\def\comecaprova{\noindent {\bf Proof:} \hspace{3mm}}
\def\terminaprova{\hfill \CaixaPreta \vspace{5mm}}
\def\mapright#1#2{\smash{\mathop{\hbox to 0.80cm{\rightarrowfill}}\limits^{#1}_{#2}}}
\def\mapleft#1#2{\smash{\mathop{\hbox to 0.80cm{\leftarrowfill}}\limits^{#1}_{#2}}}
\def\mapleftright#1#2{\smash{\mathop{\hbox to 0.80cm{\leftarrowfill \rightarrowfill}}\limits^{#1}_{#2}}}

\begin{document}

\begin{frontmatter}

\title{PHORMA: Perfectly Hashed Order Restricted Multidimensional Arrays}
\author[A1]{Lauro Lins}
\author[A2]{\hspace{-1mm}, S\'ostenes Lins}
\author[A2]{\hspace{-1mm}, S\'{\i}lvio Melo}
\address[A1]{Depto Inform\'atica da UFPE - Recife - Brazil}
\address[A2]{Depto Matem\'atica da UFPE -
Recife - Brazil}

\bibliographystyle{alpha}
\maketitle

\begin{abstract}
In this paper we propose a simple and efficient strategy to obtain
a data structure generator to accomplish a perfect hash of quite
general order restricted multidimensional arrays named {\em
phormas}. The constructor of such objects gets two parameters as
input: an $n$-vector $a$ of non negative integers and a boolean
function $B$ on the types of order restrictions on the coordinates
of the valid $n$-vectors bounded by $a$. At compiler time, the
phorma constructor builds, from the pair $a,B$, a digraph $G(a,B)$
with a single source $s$ and a single sink $t$ such that the
$st$-paths are in $1-1$ correspondence with the members of the
$B$-restricted $a$-bounded array $A(a,B)$. Besides perfectly
hashing $A(a,B)$, $G(a,B)$ is an instance of an $NW$-family. This
permits other useful computational tasks on it.

\begin{description}
\item [Keywords:] {Hash tables, Digraphs, Implicit enumeration, Nijenhuis-Wilf combinatorial families,
Constructors of objects.}

\end{description}

\end{abstract}
\end{frontmatter}

\section{Motivation and objective}

This work introduces a new type of data structure generator named
{\em phorma}, $P=(a,B)$, which consists of a positive integer
$n$-vector $a$ and a boolean function $B$ whose literals are order
restrictions on the components of the $n$-vectors $\alpha$ {\em
dominated by $a$}, that is $\alpha_i \le a_i,\ i=1,2,\ldots,n$.
The simplest example of phorma arises in the need to store a
symmetric $(p \times q)$-matrix. In this case the phorma is
$P_2^{sim}=(a=(p,q),B=\alpha_1 \ge \alpha_2)$. Our basic goal is
to enumerate in an efficient way all the equivalence classes of
indices given that the matrix is symmetric. The work that
motivates phormas, and where appears its first real use is
\cite{LinsLinsMorabito2002A}. Trying to avoid duplicates in the
huge set of of equivalences classes of indices of some
3-dimensional matrices, we were led to implement the phormas:
$P_3^{sim}=(a=(p,q,r), B = (\alpha_1 \ge \alpha_2) \ \vee \
(\alpha_2 \ge \alpha_3))$ and $P_3^{1\rightleftharpoons 2}=
(a=(p,q,r), B = (\alpha_1 \ge \alpha_2))$. The first phorma arises
when there are symmetries permuting arbitrarily all the three
coordinates. The second, when the first and second coordinates can
be interchanged, but the third is held fixed. These phormas play a
crucial role in the algorithms of \cite{LinsLinsMorabito2002A}.

To better motivate the concept and to help the reader to grasp the
definition of the general problem we treat, we discuss at length
an example of phorma (a less trivial one) arising in packing
rectangles into rectangular and $L$-shaped pieces
\cite{LinsLinsMorabito2002B}. An {\em $L$-shaped piece} is a
rectangle $R$ from which we have removed a smaller rectangle $r
\subseteq R$. Moreover $R$ and $r$ have a corner in common. By
effecting rotations, translations and reflections we may suppose
that our $L$ shaped piece has a corner in the origin and the
common vertex to $r$ and $R$ is the vertex opposite to the origin
in rectangle $R$. Positioned in this {\em canonical way}, the
$L$-piece is represented by a quadruple of real numbers
$(\alpha_1\alpha_2 \alpha_3 \alpha_4)$, with $\alpha_1 \ge
\alpha_3$ and $\alpha_2 \ge \alpha_4$, where the big rectangle $R$
has diagonal from $(0,0)$ to $(\alpha_1,\alpha_2)$ and the smaller
rectangle $r$ has diagonal from $(\alpha_3,\alpha_4)$ to
$(\alpha_1,\alpha_2)$. Let $a = a_1a_2a_3a_4$ be a positive
integer $4$-vector with $a_1 \ge a_3, a_2 \ge a_4$. In
\cite{LinsLinsMorabito2002B} we need to enumerate the canonically
positioned $L$-shaped pieces with integer coordinates $\alpha \le
a$, that is, the $L$-pieces
$\alpha=\alpha_1\alpha_2\alpha_3\alpha_4$ with (1) $\alpha_1 \ge
\alpha_3$ and (2) $\alpha_2 \ge \alpha_4$ are {\em dominated by
$a$}, $\alpha_i \le a_i,\ i=1,2,3,4$. Symmetry considerations
enable us to partition the set of $a$-bounded $L$-pieces into
equivalent classes and to distinguish a set $A$ of representatives
for these classes.

For our occupancy purposes in \cite{LinsLinsMorabito2002B} the
$L$-pieces $\alpha_1\alpha_2\alpha_3\alpha_4$ and
$\alpha_2\alpha_1\alpha_4\alpha_3$ must be considered equivalent:
one such $L$ piece is transformed into the other by a reflection
along the line passing through the origin and having slope $1$.
This is simply an axis interchange. With this in mind we have the
following order restrictions for a representative of an
equivalence class: (3) $\alpha_1 \ge \alpha_2$, otherwise we could
use $\alpha_2\alpha_1\alpha_4\alpha_3$. Also, (4) $\alpha_1 =
\alpha_2 \Rightarrow \alpha_3 \ge \alpha_4$,  otherwise we could
use $\alpha_2\alpha_1\alpha_4\alpha_3$ again. In terms of
occupancy, $\alpha_1\alpha_2\alpha_1\alpha_4$ with $\alpha_4 <
\alpha_2$, which is a degenerated $L$, can (and must) be replaced
by the rectangle $\alpha_1\alpha_2\alpha_1\alpha_2$. Analogously,
$\alpha_1\alpha_2\alpha_3\alpha_2$ with $\alpha_3 < \alpha_1$ can
be replaced by $\alpha_1\alpha_2\alpha_1\alpha_2$. In this way,
the equivalence $\alpha_1=\alpha_3 \Leftrightarrow
\alpha_2=\alpha_4$ holds. The equivalence is rewritten as two
opposite implications in the disguised form: (5) $((\alpha_1 \ne
\alpha_3) \ \vee \ (\alpha_2=\alpha_4))$ and (6) $((\alpha_2 \ne
\alpha_4) \ \vee \ (\alpha_1=\alpha_3))$. The restrictions (1) to
(6) are gathered in a boolean expression $B^L$:

\begin{center}
$B^L = (\alpha_1 \ge \alpha_3) \wedge (\alpha_2 \ge \alpha_4)\
\wedge \ (\alpha_1 \ge \alpha_2)\ \wedge \ ((\alpha_1 \ne
\alpha_2)\ \vee\
(\alpha_3 \ge \alpha_4)) \ \wedge \ $ \\
$ ((\alpha_1 \ne \alpha_3)\  \vee \ (\alpha_2=\alpha_4))\ \wedge \
((\alpha_2 \ne \alpha_4) \ \vee \ (\alpha_1=\alpha_3)).$
\end{center}

In general, a {\em phorma}, or a {\em p}erfectly {\em h}ashed {\em
o}rder {\em r}estricted {\em m}ultidimensio\-nal {\em a}rray, is a
pair $P=(a,B)$ where $a$ is an $n$-vector of positive integers and
for $\alpha$ a positive integer $n$-vector dominated by $a$, $B$
is a boolean function whose literals are of type $(\alpha_i \star
\alpha_j)$, where $\star \in \{ \le, \ge, <, >, =, \ne\}.$ The set
$A=A(P)=A(a,B)$, (of representative of the classes in the case of
the $L$-pieces), is formed by the $\alpha$'s dominated by $a$ and
satisfying $B$.

Our objective in this work is given any phorma $P=(a,B)$ to
produce a {\em constructive bijection} $h$ between $A=A(P)$ and
$\{0,1,\ldots,|A|-1\}$, so that both $h$ and $h^{-1}$ are
efficiently computable. Such functions are called {\em perfect
hash functions} \cite{Knuth1975}, \cite{CormenLeisersonRivest1990}
and their usefulness is well known in computer science.

As far as we know the problem of finding perfect hash functions
for these quite general multidimensional arrays have not been
considered before in the literature, whence the lack of more
specific references and bibliography. Our solution is based in the
theory of Nijenhuis and Wilf, chapter 13 of
\cite{NijenhuisWilf1978}. Their $NW$-combinatorial families
associates a digraph to a set of combinatorial objects in such a
way that an object is in $1-1$ correspondence with a path in the
digraph. See also a more detailed account of these combinatorial
families in  Wilf's book \cite{Wilf1990}, available at his page in
the internet. A phorma is a particular case of $NW$-combinatorial
family, specialized in boolean order specified multidimensional
arrays. Their intrinsic structure permits us to accelerate, as we
show in the final section, the calculus of $h(\alpha)$ and
$h^{-1}(w)$.

\section{The $(m,n)$-patterns}

For $m \in \{1,2,\ldots,n\}=N$, an {\em $(n,m)$-pattern
$\beta=\beta_1\beta_2\ldots\beta_n$} is a sequence of length $n$
in which each of the $m$ symbols $1,2,\ldots,m$ occurs at least
once. Given a phorma $(a,B)$ and $\alpha \in A=A(a,B)$ with
$m_\alpha$ distinct entries there exists a unique
$(n,m_\alpha)$-pattern, denoted by $\beta^\alpha$, which is order
compatible with $\alpha$: for $i \in N$, if $\alpha_i$ is the
$k$-th smallest entry among the ones appearing in $\alpha$, then
define $\beta_i^\alpha=k$. As some examples, consider the phorma
$(a=7575,B^L)$, where $B^L$ appears in the previous section. We
have $\beta^{7412}=4312$, $\beta^{5521}=3321$,
$\beta^{5533}=2211$, $\beta^{3333}=1111$. Let the set ${\mathcal
L}={\mathcal L}(P)={\mathcal L}(a,B)$ of all $(n,m)$-patterns
induced by $\alpha \in A(a,B)$,
$${\mathcal L}={\mathcal L}(P)={\mathcal L}(a,B) =
\{\beta^\alpha \ | \ \alpha \in A(a,B) \}) = (\beta^1,
\beta^2,\ldots,\beta^q),$$ be given by a list in lexicographical
order. The list ${\mathcal L}$ induces a partition of $A$ in $q$
parts: indeed, defining $[\beta^j] = \{\alpha \in A \ | \
\beta^\alpha = \beta^j \}$, we get $A = \bigcup_{j=1}^q
[\beta^j]$, with $[\beta^j] \cap [\beta^k] = \emptyset$ if $j\ne
k$. For the phorma $P^L_{7575}=(a=7575, B^L)$ we get $${\mathcal
L}^L = {\mathcal L}(P^L_{7575}) =
(1111,2121,2211,3211,3221,3321,4231,4312,4321).$$

There are only mild restrictions on the subset ${\mathcal L}$: its
cardinality, $q$, should be small enough in order for the
$\beta^j$'s to be kept in core; also, ${\mathcal L}$ should have
enough structure to be effectively generated by an implicit
enumeration scheme. In the implementation of a phorma $(a,B)$, the
first task of the {\em constructor of the data structure}
\cite{Booch1991} phorma (which is activated at compiler time) is
to obtain the list ${\mathcal L}$. How this is done? In many
applications the dimension $n$ of the phorma is small enough for
trying all $n^n$ sequences of length $n$ in symbols
$1,2,\ldots,n$, choose the ones which are $(m,n)$-patterns and
test for $B$-satisfiability \cite{CormenLeisersonRivest1990}. The
sequences that survive are added to ${\mathcal L}$. In the above
case $n^n=256$ and this simple minded approach is convenient. In
some cases it is more efficient to use appropriate
$NW$-combinatorial families \cite{NijenhuisWilf1978},
\cite{Wilf1990} which generate only $(m,n)$-patterns. Here we
avoid details of these specific families. In other cases, the list
${\mathcal L}$ is obtained by implicit enumeration. In any case,
testing $B$-satisfiability is unavoidable and is the computational
bottleneck for the phorma constructor in obtaining  ${\mathcal
L}$.

\section{The digraphs $H_\gamma$'s, $H^a$ and  $G(P)=G(a,B)$}

Throughout this work $\gamma=\gamma_1\gamma_2\ldots\gamma_m$ is an
strictly increasing $m$-sequence, $m \le n$, with entries in $N$.
For each $\alpha \in A$, let $\gamma^\alpha$ denote the strictly
increasing sequence of length $m_\alpha$ of the $m_\alpha$
distinct entries appearing in $\alpha_1 \alpha_2 \ldots \alpha_n$.
Observe that $\alpha$ is recoverable from (induced by) the pair
$(\beta^\alpha,\gamma^\alpha)$. The $\alpha$ so induced by
$(\beta,\gamma)$, where $\beta$ is an $(m,n)$-pattern and $\gamma$
is an strictly increasing $m$ sequence with entries in $N$, is
denoted $\alpha^\star(\beta,\gamma)$. As examples, in the phorma
$P_{7575}^L$, $\alpha^\star(3221,457)=7554$,
$\alpha^\star(4321,3457)=7543$, $\alpha^\star(4231,4567)=7564$. As
we shall see, the simple correspondences $\alpha \rightarrow
(\beta^\alpha, \gamma^\alpha)$ and its inverse, $(\beta,\gamma)
\rightarrow \alpha^\star(\beta,\gamma)$, inducing $\alpha
\rightleftharpoons (\beta,\gamma)$, are central for the efficient
implementation of the hash function $h$ and its inverse.

For $\beta \in {\mathcal L}$ the {\em $(a,\beta)$-maximal
increasing sequence}, denoted by $\gamma^\star(a,\beta) =
\gamma^\star_1 \gamma^\star_2 \ldots \gamma^\star_m$, is the
strictly increasing sequence of length $m$ satisfying the
following conditions: suppose that, for $1\le i \le m$, $i$ occurs
at positions $p_{i1}, \ldots, p_{ij_i}$ of $\beta$; recall that
$a=a_1a_2\ldots a_n$ and  define $\gamma^\star_m=\min \{
a_{p_{m1}}$, $a_{p_{m2}}$, $\ldots$, $a_{p_{mj_m}}\}$ and for
$i=m-1, m-2, \ldots,1$, $\gamma^\star_i$=$\min \{$ $a_{p_{i1}},
\ldots$, $a_{p_{ij_i}}$, $\gamma^\star_{i+1}-1 \}$. Observe that
$\gamma^\star(a,\beta)$ can alternatively be defined as the
lexicographically maximal $\gamma$ such that
$\alpha^\star(\beta,\gamma) \in A.$

Having constructed the list ${\mathcal L}={\mathcal L}(a,B) =
(\beta^1,\beta^2,\ldots,\beta^q)$, the next task for the phorma
constructor is to obtain a corresponding list
$\Gamma=\Gamma(a,{\mathcal L})=(\gamma^\star(a,\beta^1),$
$\gamma^\star(a,\beta^2)$,$\ldots$,$\gamma^\star(a,\beta^q))$. As
an example to help the understanding of how to obtain $\Gamma$,
consider its construction for the phorma $P_{7575}^L$. We get
$$\Gamma^L_{7575} = \Gamma(P_{7575}^L) = (5, 57, 45, 457, 457, 345,
4567,3457,3457).$$

Suppose $\gamma=\gamma_1\gamma_2\ldots\gamma_m$ is an increasing
$m$-sequence with entries in the set of positive integers. We want
to define a digraph $H_\gamma$. If $\gamma_m > m$ let
$^\leftarrow\gamma$ denote the increasing sequence of length $m$
satisfying $(^\leftarrow\gamma)_m=\gamma_m-1$ and
$(^\leftarrow\gamma)_i=\min \{
(^\leftarrow\gamma)_{i+1}-1,\gamma_i\}$, for $i=1,2,\ldots,m-1$.
If $\gamma_m=m$, then $^\leftarrow\gamma$ does not exist. If
$\gamma\ne t$, let $_{_\swarrow}\gamma$ be the sequence of length
$m-1$ obtained from $\gamma$ by removing its last entry:
$_{_\swarrow}\gamma=\gamma_1\ldots\gamma_{m-1}$. If $\gamma = t$,
then $_{_\swarrow}\gamma$ does not exit. Given $\overline{\gamma},
\widetilde{\gamma} \in \Gamma$, we say that $\overline{\gamma}
\preceq \widetilde{\gamma}$, if there is a sequence
$(\widetilde{\gamma}=\gamma^1, \gamma^2, \ldots,
\gamma^p=\overline{\gamma})$, with $\gamma^i \in \Gamma$, such
that, for each $i=1,2,\ldots,p-1$, either
$\gamma^{i+1}=^\leftarrow\hspace{-1.5mm}(\gamma^{i})$ or else
$\gamma^{i+1}=_{_\swarrow}\hspace{-1.5mm}(\gamma^{i})$. The
relation $\preceq$ is a partial order in the set $F_\infty$, of
all finite increasing sequences with integer entries. For $\gamma
\in F_\infty$, let $H_\gamma$ be the acyclic digraph whose vertex
set is $VC_\gamma = \{ \overline{\gamma} \ | \ \overline{\gamma}
\preceq \gamma\}$. The empty increasing sequence is considered a
member of $F_\infty$. It corresponds to a terminal vertex (the
unique sink), and so, is represented by $t$. From each vertex
$\overline{\gamma} \in VC_\gamma$ there are at most two outgoing
edges whose heads are $^\leftarrow\overline{\gamma}$ (if it
exists) and $ _\swarrow\overline{\gamma}$ (if it exists). These
are all the edges, what concludes the definition of $H_\gamma$. In
Fig. 1 we show all the graphs $H_{\gamma^j}$, $j=1,2,\ldots,9$,
corresponding to $\Gamma^L_{5575}$. Since
$\gamma^\star(a,\beta^4)=\gamma^\star(a,\beta^5)$ and
$\gamma^\star(a,\beta^8)=\gamma^\star(a,\beta^9)$ we get only
seven distinct digraphs. In picturing them, the direction of the
edges are implicit. They go from higher vertices to lower ones and
in the case of a draw, the direction is from right to left.

\begin{center}
\begin{tabular}{|c|} \hline \\
\includegraphics{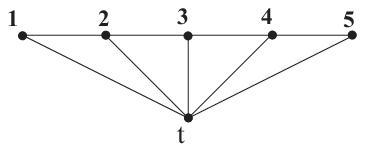}
\includegraphics{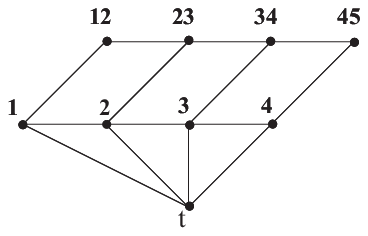}
\includegraphics{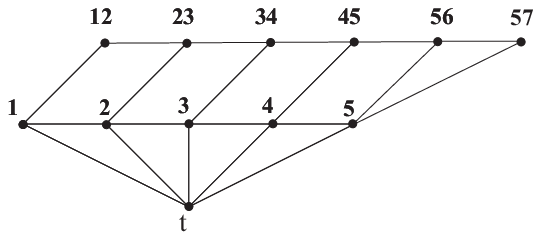}\\
\includegraphics{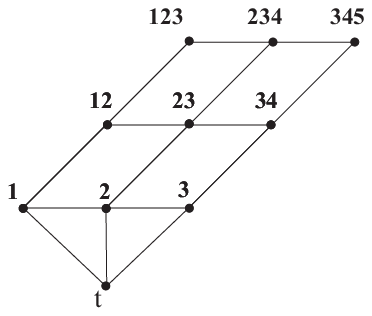} \hspace{25mm}
\includegraphics{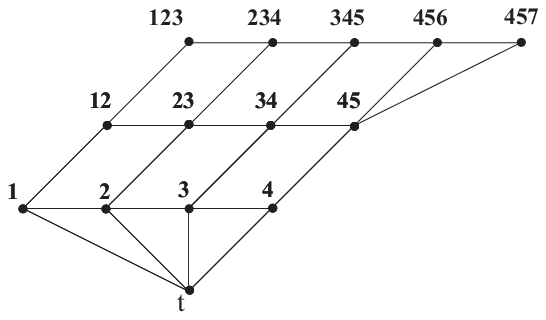}\\
\includegraphics{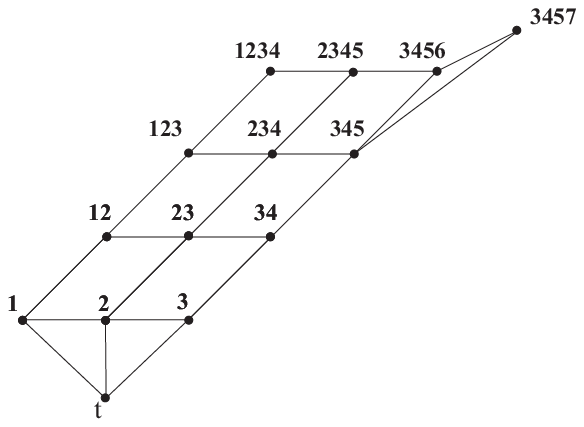}
\includegraphics{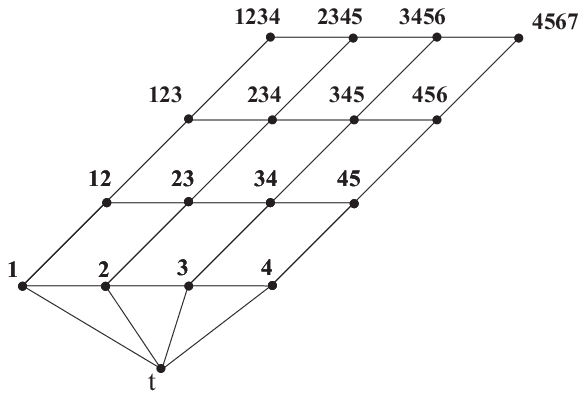}\\ \hline
Figure 1: {\sf Digraphs
$H_{5},H_{45},H_{57},H_{345},H_{457},H_{3457},H_{4567}$}
\\ \hline
\end{tabular}
\end{center}

The digraph $H^a$ is defined as the ${\mathcal L}$-indexed union
of digraphs $H_\gamma$'s:
$$H^a = \bigcup \{H_{\gamma^\star(a,\beta)} \ | \
\beta \in {\mathcal L}\}$$

The {\em digraph of the phorma}, $P=(a,B)$ is $G(a,B) = H^a
\bigcup \Lambda(a,B)$, where digraph $\Lambda(a,B)$ consists of a
root $s$ linked to vertices labelled by $\beta^j$,
$j=1,2,\ldots,q$. Each vertex $\beta^j$ is of valency $2$. The
edge from $s$ enters it and there is an edge from it to the vertex
of $H^a$ labelled by $\gamma^\star(a,\beta^j)=\gamma^j$. The total
number of edges of $\Lambda(a,B)$ is $2q$, finishing its
definition. This also concludes the definition of the digraph
$G(a,B)$. In Fig. 2 we show $G(P^L_{7575})=G^L_{7575}$. Observe
that $\Lambda(a,B)$ is depicted in dashed gray edges. The numbers
on gray are important in the computation of $h(\alpha)$ and are
explained in the next section.

\begin{center}
\begin{tabular}{|c|} \hline \\
\includegraphics{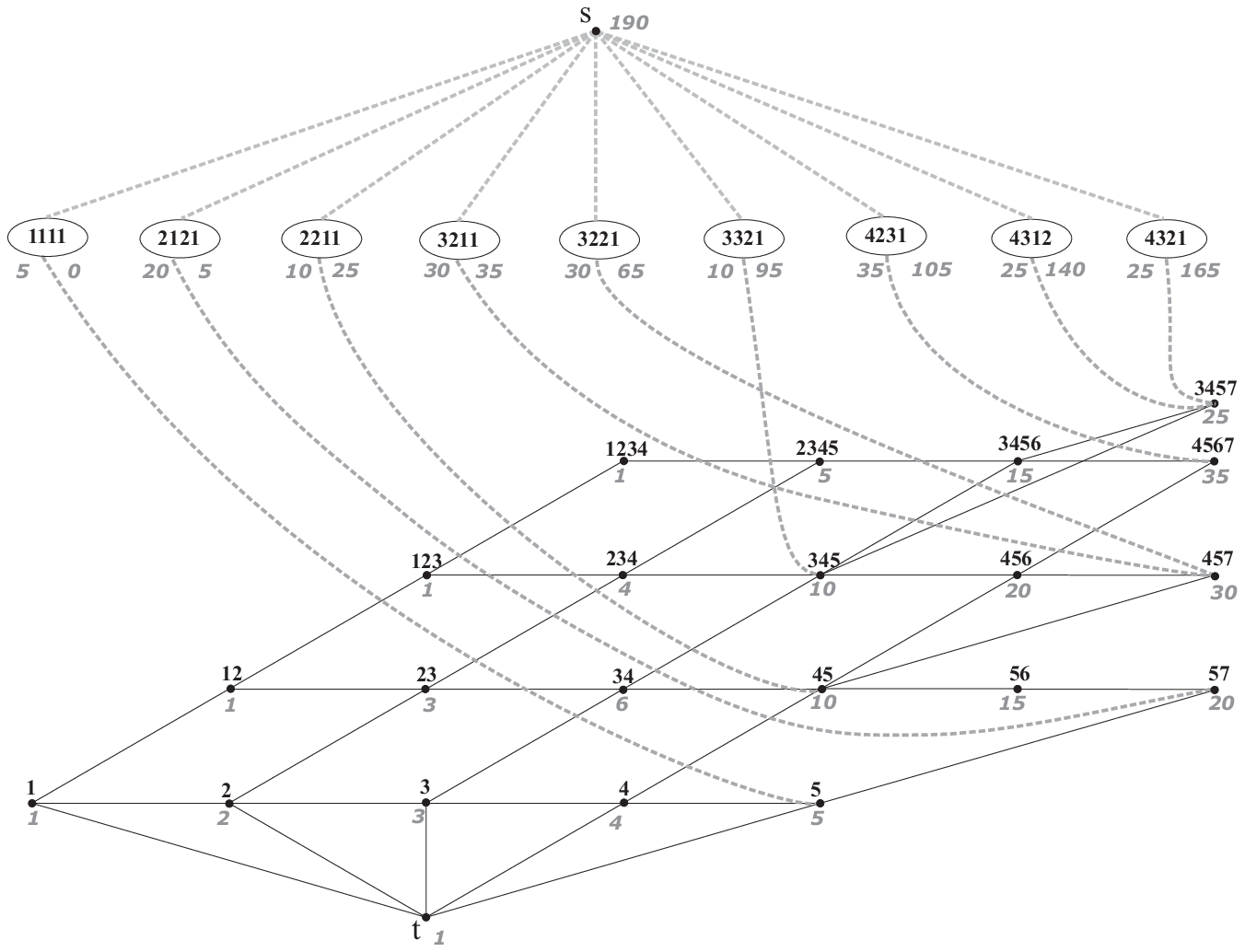}\\ \hline
Figure 2: {\sf Digraph $G^L_{7575}=G(7575,B^L)$ of the phorma
$P^L_{7575}=(7575,B^L)$}
\\ \hline
\end{tabular}
\end{center}

\section{NW-Combinatorial Families}

We briefly recall the general concept of an $NW$-combinatorial
family. An example of such an object is the digraph $H_\gamma$.
The combinatorial family that it encodes is formed by the strictly
increasing $m$-sequences $\overline{\gamma}$ with entries in $N$
dominated by $\gamma$, $\overline{\gamma}_i \le \gamma_i, i \in
N$. Also the digraph $G(a,B)$, for any phorma $(a,B)$, is an
$NW$-combinatorial family.

The following concept, introduced in \cite{NijenhuisWilf1978}, is
the central tool for this work. A {\em Nijenhuis-Wilf
combinatorial family or NW-family} is a digraph $G$, whose vertex
set is denoted by $V(G)$, having the properties below:
\begin{itemize}
\item $V(G)$ has a partial order (for $x,y \in V(G)$, $y \preceq x$ if there is a
directed path from $x$ to $y$) with a unique minimal element $t$.
For each $v \in V(G)$ the set $\{x \in V(G) \ | \ x\preceq v\}$ is
finite and includes $t$.
\item Every vertex $v$, except $t$ has a strictly positive
outvalence $\rho(v)$. For each $v \in V(G)$, the set $E(v)$ of
outgoing edges has a local rank-label $\ell_v$ , $0 \le
\ell_v(e)\le \rho(v)-1, e \in E(v)$.
\end{itemize}

Every directed path in $G$, starting from a vertex $v$ and ending
at $t$ is called a {\em combinatorial object of order $v$}. Thus,
the set of objects of order $v$ is identified with the vertex $v$.
Denote by $|v|$ the cardinality of the set of objects of type $v$,
namely, the number of paths from $v$ to $t$. In Fig. 2, the values
$|v|$ are shown as a gray number next to vertex $v$ and is the
first of the two gray numbers in the case that $v=\beta^j$. The
significance of the second gray numbers associated with the
$\beta^j$'s in Fig. 2 are explained in the final section. The
local rank-labels of the outcoming edges at $s$ in the
$NW$-combinatorial family $G(a,B)$ is given by the lexicographical
order of their heads $\beta^j$. The unique outcoming edge at
$\beta^j$ has local rank-label $0$. The local rank-labels of the
outcoming edges at a vertex $\gamma$ of $H_{\gamma^\star}$ is $0$
for the {\em west edge} (the one with head $^\leftarrow\gamma$),
if it exists, implying $1$ for the {\em southwest edge} (the one
with head $ _\swarrow\gamma$). Of course, if a vertex has only its
southwest edge, the local rank-label of this edge is $0$. Even
though in the drawings the edges arriving at $t$ are not in the
southwest direction (to decrease the width of the figures), all of
them are considered southwest edges.

From the definitions we get immediately a recursive formula for
$|v|$: $|v| = \sum \{|head(e)| \mbox{ for the edges } e \ | \
tail(e)=v\}$. This recursive formula follows from the fact that a
path from $v$ to $t$ is an outgoing edge from $v$ followed by a
path representing a combinatorial object of smaller order.
Therefore, the role of the graph $G$ defining the NW-combinatorial
family is to display how the combinatorial elements of the various
orders are inductively formed. The usefulness of the notion of
combinatorial family is that (i) a great number of usual
combinatorial objects can be encoded as paths in an NW-family;
(ii) the local rank-labels of the outcoming edges induce a {\em
unique ranking h} of the combinatorial objects of order $v$. With
respect to this ranking the following four tasks become
computationally simple and as cheap as they can be. The tasks are
exemplified and described in terms of the paths in digraph $G$,
without mentioning the specific combinatorial families that $G$
encodes. More details of the algorithms to perform these tasks can
be found in Chapter 13 of \cite{NijenhuisWilf1978}.

\noindent {\sf Task 0: counting}: What is the cardinality of the
family? {\sf Algorithm:} As we have mentioned, $|v| = \sum
\{|head(e)| \ | \ e \in E(G), tail(e)=v\}.$ It is then possible
for the constructor of the phorma to obtain the value of each
$|v|$ by recursion and to store it as an attribute of $v \in V(G)$
in a pre-processing phase (compilation time). For instance, for
the phorma $P_{7575}^L$ has cardinality $190$. This is the value
of $|s|$, in Fig. 2.

\noindent {\sf Task 1: sequencing}: Given an object in the family,
construct the ``next'' object. {\sf Algorithm:} A path starting at
$v$ and ending in $t$ is encoded by the sequence of label-ranks of
the sequence of its edges. The {\em next path} of a given path
$\pi$ in coded form is, in coded form, the lexicographic successor
of $\pi$. In coded form the $7$ paths from the vertex $v$ of the
$NW$-combinatorial family $H_{\gamma^\star}$ of Fig. 3  are: rank
$0  \rightarrow 00000$, rank $1 \rightarrow 01000$, rank $2
\rightarrow 01100$, rank $3 \rightarrow 0111$, rank $4 \rightarrow
10000$, rank $5 \rightarrow 1101$, rank $6 \rightarrow 111$. In
Theorem $1$ we shall see that these paths are in $1-1$
correspondence with the sequence of $\gamma$'s ($123, 124, 134,
234, 125, 135, 235$). This is the sequence, in rank order, of all
strictly increasing sequences of length $3$ in $\{1,2,3,4,5\}$
dominated by $\gamma^\star=235$.

\begin{center}
\begin{tabular}{|c|} \hline \\
\includegraphics{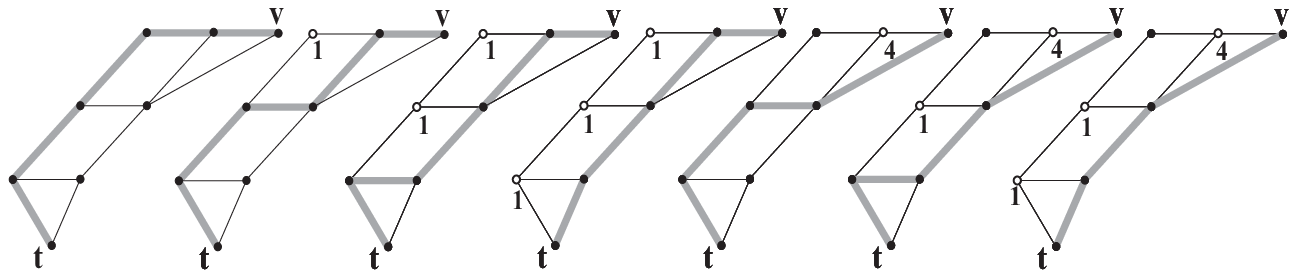}\\ \hline
Figure 3: {\sf All paths from $v=\gamma_\star=235$ to $t$ in $H_{235}$}
\\ \hline
\end{tabular}
\end{center}

\noindent {\sf Task 2: ranking (perfect hashing)}: Given an object
$\omega$ in the family, find the integer $h(\omega)$ such that
$\omega$ is the $h(\omega)$-th element in the order induced by
task 1. {\sf Algorithm:} Let an element-path $\pi$ of order $v$ of
an NW-family, $\pi=(e_1,e_2,\ldots,e_p)$ be given. The rank of
$\pi$ is defined as $h(\pi) = \sum_{i=1}^p \chi(e_i)$, where
$\chi(e)=\sum \{|head(f)| \mbox{ with } \ell_v(f) < \ell_v(e),f
\in E(v) \}$. In the $NW$-combinatorial family, this formula for
$\pi$ is particularly simple: the value $h(\pi)$ is obtained as
sum of the orders of the {\em post-falls of $\pi$} (defined in the
beginning of next section). In Fig. 3, the post-falls of the paths
are the white vertices.

\noindent{\sf Task 3: unranking}: Given an object integer $r$
construct the $r$-th member of the family. {\sf Algorithm:} Given
an integer $r$, we need to construct the $r$-th path from $v$ to
$t$. Consider $pred_v(e)$ as the highest-rank edge of the set
$\{f\in E(v)\ | \ \ell_v(f) < \ell_v(e) \}$, and let
$|head(pred_{v}(e))|=0$ if this set is empty. The required $r$-th
path's $\pi_r$ is generated as follows: $\pi_r \leftarrow
\emptyset$; $r' \leftarrow 0$; $v' \leftarrow v$; {\bf repeat}
append to $\pi_r$ the highest-rank edge $e$ of $E(v')$ such that
$r'+|head(pred_{v'}(e))| \le r$; $r' \leftarrow
r'+|head(pred_{v'}(e))|$; $v' \leftarrow head(e)$ {\bf until}
$v'=t$. It should not be difficult to check this unranking
algorithm in the paths of Fig. 3.

\noindent {\sf Task 4: getting random object}: Choose an object
uniformly at random from the given family. {\sf Algorithm:} Let
$\xi \in [0,1]$ be uniformly chosen at random; return the
$(|v|*\xi)$-th object.

\section{$1-1$ Correspondences}

Let $\pi$ be a path which starts at $\gamma^\star$ and finishes at
$s$. A {\em fall} of $\pi$ is the tail of a southwest edge, thus
$\pi$ has $m$ falls, where $m$ is the length of $\gamma^\star$. A
{\em post-fall} in $\pi$ is the vertex which is the head of an
edge whose tail is a fall. Path $\pi$ has at most $m$ post-falls.

\begin{teorema}
The $st$-paths in digraph $H_{\gamma^\star}$ are in $1-1$
correspondence with the strictly increasing $m$-sequences with
entries in $N$ which are dominated by $\gamma^\star$.
\end{teorema}
\comecaprova Any such path $\pi$ is in $1-1$ correspondence with
its sequence of {\em falls} $(\gamma^m, \dots, \gamma^2,
\gamma^1)$. Note that $\gamma^j$ ($j =1,2,\ldots,m)$, is the last
vertex of $\pi$ whose defining sequence has length $j$. Let
$\gamma^\pi_j = \gamma^j_j$, $j=1,2,\ldots,m$. Clearly
$\gamma^\pi_j = \gamma^j_j \le \gamma^\star_j$, and $\gamma^\pi$
is dominated by $\gamma^\star$. Reciprocally, given a $\gamma$
dominated by $\gamma^\star$, construct a $\pi^\gamma \equiv
(\gamma^m,\ldots,\gamma^2,\gamma^1)$ so that starting from vertex
$\gamma^\star$, the last vertex whose defining sequence has length
$j$ is $\gamma^j$ defined when we impose the equality
$\gamma^j_j=\gamma_j$. With these definitions, it follows that
$\pi^{(\gamma^\pi)}=\pi$, proving the Theorem. \terminaprova

To exemplify the above inverse constructions, consider the path
$\pi$ from $\gamma^\star=8CFJ$ to $t$ in $H_{8CFJ}$ (subscript in
base $20: A=10, B=11, \ldots, J=19$) defined by the sequence of
its falls $(8CDE,567,34,3)$. Path $\pi$ induces $\gamma^\pi$ given
by the last digits of the falls in reverse order:
$\gamma^\pi=347E$. Reciprocally, given $\gamma=347E$, starting at
$8CFJ$ the last digit of the first fall of the path $\pi^\gamma$
that we seek is the fourth digit of $\gamma$. Thus, we must go
$J-E=5$ steps to the left to arrive at $8CDE$, defining the first
fall of $\pi^\gamma$. Following the southwest edge we arrive at
$8CD$. We know that the last digit of the second fall of
$\pi^\gamma$ must be 7 (the third digit of $\gamma$). Thus we must
go $D-7=6$ steps to the left arriving at the second fall $567$. Go
southwest, arriving at $56$. The last digit of the third fall is
the second digit of $\gamma$, namely $4$. We must go $6-4=2$ steps
left arriving at the third fall $34$. Go southwest, arriving at
$3$. The last digit of the fourth fall is the first digit of
$\gamma$, namely $3$. We must go $3-3=0$ steps left to get the
fourth fall of $\pi^\gamma$, namely $3$. In this way, from
$\gamma$ and $\gamma^\star$ we have obtained the sequence of falls
$(8CDE, 567, 34, 3)$. This sequence of falls define $\pi^\gamma$.
Clearly, $\pi^{(\gamma^\pi)}=\pi$.

\begin{center}
\begin{tabular}{|c|} \hline \\
\hspace{-1mm}\includegraphics{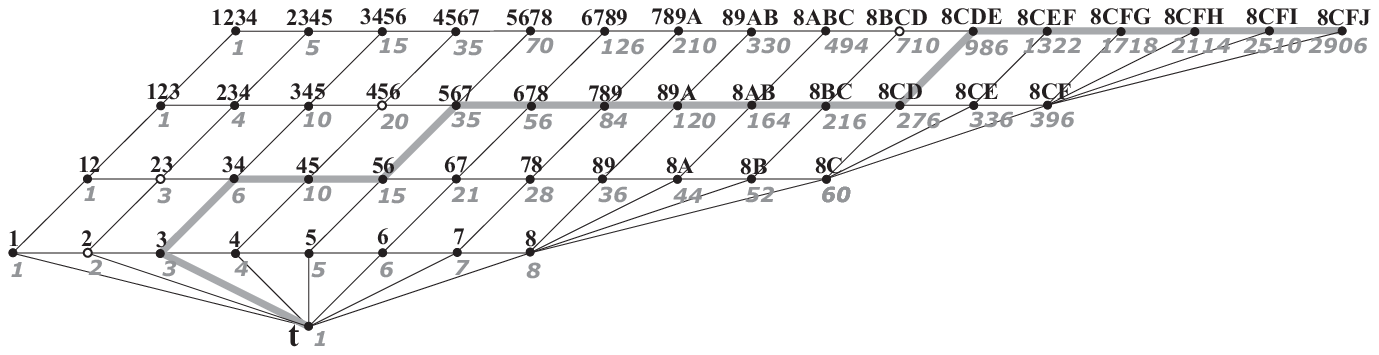}\\ \hline Figure 4: {\sf
Digraph $H_{8,12,15,19}$ (vertex labels in base 20)}
\\ \hline
\end{tabular}
\end{center}

\begin{teorema}
For any phorma $P=(a,B)$, the $st$-paths in digraph $G(a,B)$ are
in $1-1$ correspondence with $A(a,B)$.
\end{teorema}
\comecaprova Let $\pi$ be an $st$-path in digraph $G(a,B)$ and
$(s,\beta^j,\gamma^1,\gamma^2,\ldots,\gamma^p,t)$ be the sequence
of vertices in $\pi$. By Theorem $1$, the subpath from $\gamma^1$
to $t$ which is in $H_{\gamma^1}$ defines a $\gamma^\pi$ dominated
by $\gamma^1$. The $\alpha$ which corresponds to $\pi$ is
$\alpha^\star(\beta^j,\gamma^\pi)$. Reciprocally, given $\alpha
\in A(a,B)$, consider the pair $(\beta^\alpha,\gamma^\alpha)$. Let
$\pi'$ be the path in $H_{\gamma^\star(a,\beta^\alpha)}$ from
$\gamma^\star(a,\beta^\alpha)$ to $t$ which corresponds to
$\beta^\alpha$, given by Theorem 1. The $st$-path $\pi$ in
$G(a,B)$ which corresponds to $\alpha$ is obtained from $\pi'$ by
pre-fixing to it the two edges, from $s$ to $\beta^\alpha$ and
from $\beta^\alpha$ to $\gamma^\star(a,\beta^\alpha)$. The
correspondences $\pi \mapsto \alpha$ and $\alpha \mapsto \pi$ are
inverses establishing the Theorem. \terminaprova

\section{Implementation Issues}

{\bf Entering a generic phorma type boolean function $B$.} A
convenient way to store such boolean functions is by means of a
tree $T(B)$ with three types of internal nodes:
$\vee$-nodes,$\wedge$-nodes, $\neg-$nodes. The leaves of the tree
correspond to the basic constituent boolean functions of type
$\alpha_i \star \alpha_j$, where $\star \in
\{\le,\ge,<,>,=,\ne\}$. The $\neg-$nodes (negation operator) must
have at most one child. Note that each subtree rooted at an
internal $\diamond$-node $v$ $(\diamond \in \{\vee, \wedge,
\neg\})$  is a boolean tree obtained by taking the
$\diamond$-operation of the boolean tree(s) corresponding to the
children of $v$.  Given an $\alpha$ it is rather easy to decide
$B$-satisfiability of $\alpha$, by evaluating from the leaves up
and arriving to the root of $T(B)$.

{\bf Properly storing $H^a$.} Let $a^\star$ be the maximum of the
$a_i$'s. Consider a bidimensional array $R[0..n,1..a^\star]$, in
which cell $R[m,p]$ contains the address of a simple linked list
containing in $\gamma$-lexicographical order all the pairs
$(_\swarrow \gamma,|\gamma|)$ in which $\gamma$ is a vertex of
$H^a$, and, as a sequence, has length $m$ and satisfies
$\gamma_m=p$. The need to use the pairs $(_\swarrow
\gamma,|\gamma|)$ become clear to efficiently perform the rank
operation, as explained below. The maximum length of the list
$R[m,p]$, denoted by $|R[m,p]|$ is $|{\mathcal L}(a,B)|$, however,
these lengths tends to be very small. In particular, if all the
entries of $a$ are equal, or if $p \le 2$, $|R[m,p]|=1$. In the
example of Fig. 2, the only entry $|R[m,p]|$ which is not $1$ is
$|R[4,7]|=2$. Graph $H^a$ is stored as a hash table
$R[0..m,1..a^\star]$ in which the pairs $(_\swarrow
\gamma,|\gamma|)$ having $\gamma$ with the same length $m$ and the
same last element $p$ are stored together in a
$\gamma$-lexicographically ordered list (to resolve the
conflicts). We consider that a binary search in the list $R[m,p]$
to locate the specific pair $(_\swarrow \gamma,|\gamma|)$ is good
enough.

{\bf Ranking in the NW-Family $H_{\gamma^\star}$}. In order to
obtain the rank of $\gamma \in H_{\gamma^\star}$ of length $m$
based in a usual pointer implementation
(\cite{CormenLeisersonRivest1990}) of $H^a$ we may need to walk
along a path $\pi_\gamma$ of length $a^\star+m$, where $a^\star =
\max \{a_i \ | \ i = 1,2,\ldots,n \}$. By using the above hash
table to store $H^a$, we do the job in $m$ steps. This is a
critical speeding up improvement, since in most applications
$a^\star >> m$. Let $h_{\gamma^\star}(\gamma)$ denote the rank of
$\gamma$ in $H_{\gamma^\star}$. Let $m$ be the length of $\gamma$
and $\gamma^m, \ldots, \gamma^2, \gamma^1$ the sequence of falls
of $\pi^\gamma$. We know that $h_{\gamma^\star}(\gamma)$ is the
sum of the orders of the corresponding post-falls,
$|^\leftarrow\gamma^m|+ |^\leftarrow\gamma^{m-1}|+\ldots+
|^\leftarrow\gamma^2|+|^\leftarrow\gamma^1|$. In this rank
formula, if $^\leftarrow\gamma^i$ does not exists then
$|^\leftarrow\gamma^i|$ is defined as $0$. Let
$^{\leftarrow\leftarrow}\gamma$ be denoted by
$^{2\leftarrow}\gamma$, $^{\leftarrow\leftarrow\leftarrow}\gamma$
be denoted by $^{3\leftarrow}\gamma$, etc; also
$^{0\leftarrow}\gamma = \gamma$. If $\gamma_m \ge m+j$, then
$^{j\leftarrow}\gamma$ exists and is given by
$(^{j\leftarrow}\gamma)_i= \min \{\gamma_m-j-m+i,\gamma_i\}$, for
$i=1,2,\ldots,m.$ The $\gamma^i$'s can be found as follows: Let
$\xi_m=\gamma^\star_m-\gamma_m$ and
$\gamma^m=\hspace{1mm}^{\xi_m\leftarrow}\hspace{-1mm}(\gamma^\star)$.
For $i=m-1,m-2,\ldots,2,1$ let $\xi_i=\gamma^{i+1}_i - \gamma_i$
and $\gamma^i = \hspace{1mm}
^{\xi_i\leftarrow}(_\swarrow\gamma^{i+1})$. Since we need only
$|^\leftarrow\gamma^m|, |^\leftarrow\gamma^{m-1}|, \ldots,
|^\leftarrow\gamma^2|,|^\leftarrow\gamma^1|$, it is enough to
store, for each vertex $\gamma \in V(H^a)$, the pair of entries
$(\gamma, |\gamma|)$. All such pairs with $\gamma$ of length $m$
and $\gamma_m=p$ are stored in the list $R[m,p]$, ordered
lexicographically by $\gamma$. Since all $\gamma$'s in the pairs
$(\gamma,|\gamma|)$'s stored at $R_m(p)$ satisfy $\gamma_m=p$, we
may drop the last entry of $\gamma$ and store the pairs
$(_{_\swarrow}\gamma,|\gamma|)$.

{\bf Ranking in the NW-Family $G(A,L)$}. For a vertex $v$ with at
most one incoming edge $e_v$ of an NW-family, let $||v|| =
\chi(e_v)$. This is the case of $s$ and of $\beta$ in ${\mathcal
L}(a,B)$. The value of $||s||$ is zero and the values of
$||\beta||$'s are pre-computed for each $\beta$. Translating from
the general recipe for ranking in an NW-family to our specific
case,
$$h(\alpha) = ||s|| + ||\beta^\alpha|| +
h_{\gamma^\star(a,\beta^\alpha)}(\gamma^\alpha) = ||\beta^\alpha||
+ h_{\gamma^\star(a,\beta^\alpha)}(\gamma^\alpha).$$
$\beta^\alpha$ is located by a binary search on ${\mathcal
L}(a,B)$. From the pair $(a,\beta^\alpha)$ we compute
${\gamma^\star(a,\beta^\alpha)}$ which is the entry point in $H^a$
of the path $\pi^\alpha$.

\section{Conclusion} We have defined a data structure generator
which permits the perfect hash of order restricted
multidimensional arrays $A(a,B)$. The restrictions accord a very
general type of boolean functions $B$ formed by order restricting
pairs of entries of the array in arbitrary ways. Our scheme is the
conjunction of two ideas: (1) To make a list ${\mathcal L}(a,B)$
of the order patterns which induce a partition of $A(a,B)$. (2)
Distinct patterns which have the same increasing sequences of
symbols are treated together. In consequence, an $n$-vector
$\alpha \in A(a,B)$ is subdivided into two pieces of information
$\beta$, the pattern associated to $\alpha$ and  $\gamma)$, the
increasing sequence of distinct symbols appearing in $\alpha$.
This encoding has the power of perfectly addressing huge arrays
$A(a,B)$ by means of logarithmically smaller digraphs $G(a,B)$
($NW$-combinatorial families). This general type of perfect hash
scheme does not seem to have been treated before in the
literature. In particular, its applications to database systems is
a possible source of relevant applications and remains to be
investigated.


\begin{thebibliography}{99}

\bibitem{Booch1991} G. Booch. Object oriented design with
applications. {\em The Benjamim/Cummings Publshing Coompany, Inc,
ISBN 0-8053-0091-0, 1991}

\bibitem{CormenLeisersonRivest1990} T. Cormen, C. Leiserson and R. Rivest.
Introduction to Algorithms. {\em The MIT Electrical Engineering
and Computer Science Series --- The MIT Press, McGraw-Hill Book
Company, ISBN 0-262-03141-8, 1990.}


\bibitem{Knuth1975} D. Knuth, The art of computer programming, vol
3. {\em Adinson Wesley (second edition) 1975}.

\bibitem{LinsLinsMorabito2002A} L. Lins, S. Lins and R. Morabito.
An $n$-tet graph approach to non-guillotine packings of
$n$-dimensional boxes into an $n$-container. {\em European Journal
of Operations Research 141 (2002) 421-439}.

\bibitem{LinsLinsMorabito2002B} L. Lins, S. Lins and R. Morabito.
An $L$-approach for packing $(\ell,w)$-rectangles into rectangular
and $L$-shaped pieces. {\em Submitted to the Journal Operations
Research Society (August 2002)}.

\bibitem{NijenhuisWilf1978} A. Nijenhuis and H. S. Wilf. Combinatorial algorithms
for computers and calculators. {\em Academic Press (second
edition) 1978}.

\bibitem{Wilf1990} H. S. Wilf. East side, west side $\ldots$
Available in PDF at the home page of the author:
www.cis.upenn.edu/$\sim$wilf, 1990.


\end{thebibliography}
\end{document}